\documentclass[]{elsart}
\usepackage{graphicx}
\usepackage{epsfig}
\usepackage{amssymb}
\usepackage{dcolumn}
\usepackage{bm}

\begin{document}

\begin{frontmatter}

\title{Effects of Boron Purity, Mg Stoichiometry and Carbon Substitution on Properties of Polycrystalline MgB$_{2}$}

\author{R. A. Ribeiro}, \author{S. L. Bud'ko}, \author{C. Petrovic}, \author{P. C.
Canfield}

\address{Ames Laboratory and Department of Physics and Astronomy\\
Iowa State University, Ames, IA 50011 USA}

\begin{abstract}
By synthesizing MgB$_{2}$ using boron of different nominal purity
we found values of the residual resistivity ratio ($RRR = R(300 K)
/ R(42 K)$) from 4 to 20, which covers almost all values found in
literature. To obtain high values of $RRR$, high purity reagents
are necessary. With the isotopically pure boron we obtained the
highest $RRR \sim$ 20 for the stoichiometric compound. We also
investigated Mg$_{x}$$^{11}$B$_{2}$ samples with 0.8 $< x <$ 1.2.
For the range Mg$_{0.8}$$^{11}$B$_{2}$ up to
Mg$_{1.2}$$^{11}$B$_{2}$ we found average values of $RRR$ between
14 and 24. For smaller variations in stoichiometry ($x=1\pm 0.1$)
$RRR =  18 \pm 3$. All of our data point to the conclusion that
high $RRR$ ($\sim 20$) and low $\rho_{0}$ ($\leq 0.4 \mu \Omega
cm$) are intrinsic material properties associated with high purity
MgB$_{2}$. In addition we have performed initial work on
optimizing the formation of carbon doped MgB$_{2}$ via the use of
B$_{4}$C. Nearly single phase material can be formed by reaction
of nominal Mg(B$_{0.8}$C$_{0.2}$)$_{2}$ for 24 hours at
$1200^{\circ}C$.  The $T_{c}$ for this composition is between
$21.9 K$ and $22.7 K$ (depending on criterion).
\end{abstract}

\begin{keyword}
MgB$_{2}$ \sep stoichiometry \sep transport properties \PACS
74.70.Ad \sep 74.25.Fy
\end{keyword}
\end{frontmatter}

\section{Introduction}\label{sec:intro}

Since the discovery of superconductivity in the compound MgB$_2$
by Akimitsu and co-workers \cite{sendai,akimitsu}, considerable
progress has been made in the understanding of the fundamental
properties of this material. Within weeks of the announcement of
this discovery, it was established that high purity, very low
residual resistivity samples of MgB$_{2}$ could be synthesized by
exposing boron powder or filaments to Mg vapor at temperatures at
or near $950^{\circ}C$ for as little as two hours
\cite{budkoPRL,finnemore01,canfield01}. Samples with residual
resistivity ratio [$RRR = R(300 K)/R(42 K)$] values in excess of
20 and residual resistivities as low as 0.4 $\mu \Omega cm$ were
synthesized by this method. Such a low resistivity in an
intermetallic compound with a superconducting critical
temperature, $T_{c}$, near $40 K$ was of profound physical, as
well as engineering, interest. The implications of this high $RRR$
and low $\rho_0$ ranged from large magneto-resistances to
questions of how a material with such an apparently large
electron-phonon coupling could have such a small normal state
resistivity. On the applied side, a normal state resistivity of
0.4 $\mu \Omega cm$ for temperatures just above $T_{c}$ means that
MgB$_{2}$ wires would be able to handle a quench with much greater
ease than, for example, Nb$_{3}$Sn wires which have a $\rho_0$
that is over an order of magnitude larger for $T \sim 20 K$
\cite{canfield01}. Unfortunately, other techniques of synthesizing
MgB$_{2}$ had difficulties in achieving such high $RRR$ or low
$\rho_0$ values \cite{chen,lee,pradhan,kijoon,xue}. In some cases,
the authors of these papers have concluded that the resistivity of
their samples must be the intrinsic resistivity and that higher
$RRR$ values or lower residual resistivity values must somehow be
extrinsic. In order to address these concerns and in order to shed
some light on how low resistivity samples can be grown, we have
studied the effects of boron purity and magnesium stoichiometry on
sintered pellet samples. Based on these measurements, we conclude
that the purity of the boron used to make the MgB$_{2}$ is a
dominant factor in determining the ultimate, low temperature,
normal state resistivity of the sample, and that $RRR$ values as
high as 20 and residual resistivities as low as 0.4 $\mu \Omega
cm$ are intrinsic materials properties of high purity MgB$_{2}$
\cite{ribeiro}.

Once reproducible synthesis of high purity, single phase,
MgB$_{2}$ is understood and believed to be somewhat controllable,
the next question to be addressed is: can  MgB$_{2}$ be doped in a
reliable manner?  Whereas the effects of Al substitution for Mg
were addressed very early on \cite{slusky,lorenz,pena},
substitutions on the boron site have been somewhat more difficult.
There have been some attempts to substitute carbon for boron
\cite{paranthaman,ahn,takenobu,mickelson} with results varying
from very little effect on $T_{c}$ \cite{paranthaman} to shift of
$T_{c}$ down to $\sim 35 K$ \cite{ahn,takenobu} or $32 K$
\cite{mickelson} depending upon the nominal carbon concentration
as well as the sample synthesis route.  By far the most appealing
route to producing carbon doped MgB$_{2}$ is to start with the
carbon mixed with the boron on the length scale of a unit cell.
Starting with a boron rich compound such as B$_{4}$C appears to
offer just such a route \cite{mickelson}. In order to determine
how to synthesize as close to a single phase sample as possible we
have measured powder X-ray diffraction spectra as well as
temperature dependent magnetic susceptibility and electrical
transport on a series of samples synthesized at different
temperatures.

\section{Sample synthesis}\label{sec:synthesis}

Samples of MgB$_{2}$ for this study were made in the form of
sintered pellets. The sintered pellets were made by sealing
stoichiometric amounts of Mg($99.9\%$) and B into Ta tubes,
placing these tubes (sealed in quartz) into furnaces heated to
$950^{\circ}C$ for 3 hours, and then quenched to room temperature
\cite{budkoPRL}. For the initial studies of the effects of boron
purity, stoichiometric MgB$_{2}$ was synthesized and the quality
of the boron was varied. For the studies of magnesium
stoichiometry, nominal stoichiometries that ranged from
Mg$_{0.8}$$^{11}$B$_{2}$ to Mg$_{1.2}$$^{11}$B$_{2}$ were used and
samples were synthesized with 99.95\% pure, isotopically enriched
$^{11}$B \cite{ribeiro}.

Given that our synthesis technique involves a reaction between
solid boron and Mg vapor we felt that any attempt at boron site
substitution required that the dopant and the boron be mixed at an
atomic level before exposure to Mg vapor. The compound B$_{4}$C is
ideal given that it is boron rich and includes carbon - a likely
dopant. Earlier work \cite{mickelson} indicated this may be a
viable route, but also appeared to be somewhat preliminary. The
carbon doped MgB$_2$ sintered pellets were made by sealing lumps
of Mg ($99.9\%$) and B$_4$C ($99\%$ - Alfa Aesar) into Ta tubes,
following same procedure for the pure MgB$_{2}$, and heating to a
variety of temperatures. For this work, the samples were heated
for 24 hours to four different temperatures: ($750^{\circ}C$,
$950^{\circ}C$, $1100^{\circ}C$, $1200^{\circ}C$). In order to
compare our samples with the initial studies using B$_4$C as
starting reagent (Mickelson \textit{et al.} \cite{mickelson}), we
also prepared samples by heating to $600^{\circ}C$ for 2 hours and
then $700^{\circ}C$ for 2 more hours. The initial stoichiometric
ratio in all these procedures was 5:2 of Mg:B$_4$C giving a
nominal composition of Mg(B$_{0.8}$C$_{0.2}$)$_{2}$.

A.C. electrical resistance measurements were made using Quantum
Design MPMS and PPMS units. Platinum wires for standard four-probe
configuration were connected to the samples with Epotek H20E
silver epoxy. LR 400 and LR 700 A.C. resistance bridges were used
to measure the resistance when the MPMS units were used to provide
the temperature environment. Powder X-ray diffraction measurements
were made using a Cu $K_{\alpha}$ radiation in a Scintag
diffractometer and a Si standard was used for all runs. The Si
lines have been removed from the X-ray diffraction data, leading
to apparent gaps in the powder X-ray spectra.

\section{Effects of Boron Purity}\label{sec:boron}

Figure 1 presents powder X-ray diffraction spectra for three
samples with varying nominal boron purities: 90\% purity, 99.99\%
purity, and the 99.95\% purity, isotopically pure $^{11}$B. By
comparing the two upper panels to the bottom panel it can be seen
that the strongest MgB$_{2}$ lines are present in all three
samples. The spectrum shown in the upper panel, the data taken on
the sample made from boron with only a 90\% nominal purity, also
has weak Mg and MgO lines present. This is not inconsistent with
the fact that the primary impurity in the 90\% boron is associated
with Mg.

Figure 2 displays the normalized resistance, $R(T)/R(300 K)$, of
MgB$_{2}$ pellets that were made using the five different types of
boron powderas described in Table 1. Each curve is the average of
three resistance curves taken on different pieces broken off of
each pellet. Figure 2 demonstrates that $RRR$ values can range
from as low as 4 to as high as 20 depending upon what source of
boron is used. Among the natural boron samples examined there is a
steady increase in $RRR$ as the purity of the source boron is
improved. The MgB$_{2}$ synthesized from the isotopically pure
boron appears to have the best $RRR$, although it's nominal purity
is somewhat less than that of the 99.99\% pure natural boron, but
those skilled in the art will realize that claims of purity from
different companies can vary dramatically. In addition, it is
likely that the isotopically pure boron was prepared in a somewhat
different manner from the other boron powders used (very likely
using a boron fluoride or boric acid or any of its complexes as an
intermediate phase, in order to achieve isotopic separation). The
primary point that figure 2 establishes is that the purity of the
boron used can make a profound difference on the normal state
transport properties.

In Fig. 3 the same resistance data is plotted, but instead of
simply normalizing the data at room temperature the data is
normalized to the temperature derivative at room temperature. This
is done to see if the resistance curves differ only by a
temperature independent residual resistivity term: i.e. this
normalization is based upon the assumption that the slope of the
temperature dependent resistivity at room temperature should be
dominated by phonon scattering, and therefore be the same for each
of these samples. As can be seen, this seems to be the case, at
least to the first order. By using higher purity boron we are able
to diminish the additive, residual resistance by a factor of
approximately five.

The insets to figs. 2 and 3 indicate that there is a monotonic
improvement in $T_{c}$ as the boron purity (or $RRR$ value) is
increased. $T_{c}$ values vary from just below $38 K$ to just
above $39 K$ depending upon which boron is used. It should be
noted that similar behavior have been seen in other
polycrystalline samples with poor $RRR$ values
\cite{chen,lee,pradhan,kijoon,xue}.

Based upon these results, we choose the isotopically pure $^{11}$B
for the further study of the effects of Mg stoichiometry on
MgB$_{2}$ pellet samples. But before we proceed to the next
section, it is worth noting that one of the difficulties
associated with the samples made by other research groups may well
be due to the use of boron with less than the highest purity. In
addition, to our knowledge very few other groups have been using
the Eagle-Picher isotopically pure boron in the samples for
electrical transport measurements. It should be noted though that
recent measurements\cite{larb} have reproduced these results using
Eagle-Picher $^{11}$B.

\section{Effects of Magnesium Stoichiometry}\label{sec:stoichiometry}

In order to study the effect of magnesium stoichiometry on the
transport properties of Mg$^{11}$B$_{2}$, a series of
Mg$_{x}$$^{11}$B$_{2}$ ($0.8 \leq x \leq 1.2$) samples were
synthesized. Figure 4 presents powder X-ray diffraction patterns
for the extreme members of the series (top and bottom panels) as
well as for the stoichiometric Mg$^{11}$B$_{2}$ (middle panel). In
all cases the lines associated with the Mg$^{11}$B$_{2}$ phase are
present. For the Mg$_{0.8}$$^{11}$B$_{2}$ sample, there is a weak
line seen at $2\theta = 35.8^{\circ}$ that is associated with
MgB$_{4}$ (marked with a +). This is consistent with the fact that
there was insufficient Mg present to form single phase
Mg$^{11}$B$_{2}$. For the Mg$_{1.2}$$^{11}$B$_{2}$ sample there
are strong diffraction lines associated with Mg (marked with *).
This too is consistent with the stoichiometry of the sample:
Mg$^{11}$B$_{2}$ is the most Mg-rich member of the binary phase
diagram, therefore any excess Mg will end up as unreacted Mg. The
X-ray diffraction pattern for the stoichiometric Mg$^{11}$B$_{2}$
shows much smaller peaks associated with a small amount of both
MgB$_{4}$ and Mg phases. This pattern is different from the one
shown in Fig. 2 in that this sample was reacted for 3 hours,
whereas the sample used in Fig. 1 was reacted for 4 hours. Given
that all of the samples used for the Mg-stoichiometry study were
reacted for 3 hours, it is appropriate to show this powder
diffraction set along with the other members of the series. It
should be noted that there is continuous change in the nature of
the second phases in the samples. For Mg deficient samples there
is only MgB$_{4}$ as a second phase. For the stoichiometric
Mg$^{11}$B$_{2}$ samples there are either no second phases or very
small amounts of both MgB$_{4}$ and Mg (depending upon reaction
times), and for the excess Mg samples there is no MgB$_{4}$, but
clear evidence of excess Mg.

Figure 5 presents normalized resistance data for five
representative Mg$_{x}$$^{11}$B$_{2}$ pellets. In each case the
curve plotted is the average of three or more samples cut from the
same pellet. There is far less variation between the different
pellets in this case than there was for the case of boron purity
(Fig. 2). This is most clearly illustrated by the fact that the
values of the $R(T)$ collapse almost completely onto a single
manifold as viewed on full scale. Figure 6 plots the $RRR$ values
for each of the individual samples (shown as the smaller symbols)
as well as the $RRR$ of the average curve. As can be seen, the
$RRR$ values increase slowly from $\sim 14$ for
Mg$_{0.8}$$^{11}$B$_{2}$ to $\sim 18$ for Mg$^{11}$B$_{2}$. This
is followed by an increase in $RRR$ values for excess Mg, with
Mg$_{1.2}$$^{11}$B$_{2}$ having an $RRR$ value of $\sim 24$. The
important point to note is that even, for the most Mg deficient
sample, the lowest measured $RRR$ value is significantly greater
than 10. At no point in this series we find samples with $RRR$
values of 3, 6, or 10, even when a clear MgB$_{4}$ second phase is
present. For samples ranging from Mg$_{0.9}$$^{11}$B$_{2}$ to
Mg$_{1.1}$$^{11}$B$_{2}$ (dotted box in figure 6) the average
$RRR$ values cluster around $RRR = 18 \pm 3$. These data indicate
that, for sintered pellets, $RRR$ values of 18 can be associated
with stoichiometric Mg$^{11}$B$_{2}$ in pellet form.

Whereas the effects of excess Mg are relatively minor in these
samples (given their low intrinsic resistivities), these effects
can still be clearly seen. In addition to the increase in the
$RRR$ value, there is a change in the form of the temperature
dependence of the resistance. This can be best seen in Fig. 7, in
which the resistance data have been normalized to its room
temperature slope. The data for all $x$ values less than 1.0 are
similar (to within small differences in residual resistivity) and
can be collapsed onto a single curve. On the other hand, the
resistance data for the $x=1.1$ and $x=1.2$ are qualitatively
different. They start out with somewhat higher normalized
resistance data than the stoichiometric sample, and then below
$100 K$ cross below the stoichiometric sample. This is shown in
Fig. 7 by representing the data for $x=1.2$ as a dashed line and
can also be seen in the inset (stars). This change in behavior is
very likely due to the increasing effects of having Mg in parallel
(and series) with the MgB$_{2}$ grains. As can be seen in Fig. 7,
this effect becomes larger as the amount of excess Mg is
increased. This deviation from the MgB$_{2}$ resistance curve may
actually serve as a diagnostic for the detection of excess Mg.

For further details about the effects of boron purity, Mg
stoichiometry or the results of our studies on MgB$_{2}$ wires
segments see reference \cite{ribeiro}.

\section{Carbon doping} \label{sec:carbon}

 Figure 8 presents the powder X-ray diffraction patterns for nominal
Mg(B$_{0.8}$C$_{0.2}$)$_{2}$ samples reacted for 24 hours at
temperature of 750, 950, 1100 and $1200^{\circ}C$.  The top
pattern is from a sample that was reacted for 2 hours at
$600^{\circ}C$ and then reacted for 2 more hours at
$700^{\circ}C$.  This sample was made using the temperature / time
schedule outlined in reference \cite{mickelson} and serves as a
point of comparison.  There is a clear decrease in the signal
coming from impurity phases in the material as the reaction
temperature is increased.  The reactions carried out at either
1100 or $1200^{\circ}C$ appear to be approaching single phase.

The temperature dependent magnetic susceptibility and electrical
resistance for these samples are presented in Figs. 9 and 10
respectively.  As the phase purity of the material is improving
the onset temperature (see insets) decreases and the transition
sharpens.  The sample synthesized at $1200^{\circ}C$ for 24 hours
has $T_{c} = 22.7 K$ based on zero resistance criterion and $T_{c}
= 21.9 K$ based on an onset of diamagnetism criterion. It should
be noted that, based on the magnetization data, for the sample
synthesized at $950^{\circ}C$ the majority of the sample appears
to have a $T_{c} \sim 16 K$, well below the onset temperature of
$23.5 K$. This, as well as the other transition temperatures found
for samples reacted at $750^{\circ}C$ or lower temperature may
well be associated with inhomogeneities in these samples and may
indicate that even further carbon doping levels are possible. In
addition, there is still a slight increase in $T_{c}$ when the
reaction temperature is increased from $1100^{\circ}C$ to
$1200^{\circ}C$. This may indicate that a further increase in
reaction time or temperature will yield some further (slight)
increase in $T_{c}$. These data represent a first step in the
optimization and study of carbon doped MgB$_{2}$ with the nominal
composition of Mg(B$_{0.8}$C$_{0.2}$)$_{2}$.  A more detailed
study is part of an ongoing research project.

\section{Conclusion} \label{sec:conclusion}

In summary, through the synthesis of various MgB$_{2}$ samples
with different nominal boron purities we found values of $RRR$
from 4 to 20, which covers almost all values found in literature.
To obtain high values of $RRR$, high purity reagents are
necessary. With the isotopically pure boron we obtained the
highest $RRR \sim$ 20 for the stoichiometric compound. We also
investigated Mg$_{x}$$^{11}$B$_{2}$ samples with 0.8 $< x <$ 1.2.
These have shown that from the most Mg deficient samples we
observe inclusions of the MgB$_{4}$ phase, and no evidence of Mg.
For samples with excess Mg we do not observe any MgB$_{4}$. For
the range Mg$_{0.8}$$^{11}$B$_{2}$ up to Mg$_{1.2}$$^{11}$B$_{2}$
we found average values of $RRR$ between 14 and 24. For smaller
variations in stoichiometry ($x=1\pm 0.1$) $RRR =  18 \pm 3$. All
of our data point to the conclusion that high $RRR$ ($\geq 20$)
and low $\rho_{0}$ ($\leq 0.4 \mu \Omega cm$) are intrinsic
materials properties associated with high purity MgB$_{2}$
\cite{ribeiro}.

Our initial work on optimizing the formation of carbon doped
MgB$_{2}$ via the use of B$_{4}$C indicates that nearly single
phase material can be formed by reaction of nominal
Mg(B$_{0.8}$C$_{0.2}$)$_{2}$ for 24 hours at $1200^{\circ}C$.  The
$T_{c}$ for this composition is between $21.9 K$ and $22.7 K$
(depending on criterion for $T_{c}$ used). Further work on the
optimization and characterization of this compound is ongoing, but
it appears that Mg(B$_{0.8}$C$_{0.2}$)$_{2}$ may offer a very
useful window on the rather novel physics associated with
MgB$_{2}$.

{\bf Acknowledgements}

Ames Laboratory is operated for the US Department of Energy by
Iowa State University under Contract No. W-7405-Eng-82. This work
was supported by the Director for Energy Research, Office of Basic
Energy Sciences. The authors would like to thank N. Kelson for
drawing our attention to earlier research on the properties of
MgB$_2$ \cite{kost}, M. A. Avila, D. K. Finnemore and N.E.
Anderson, Jr. for helpful assistance and many fruitful
discussions.

\newpage
{\bf Table}

Table 1: Boron form and purity (as provided by the seller).

\newpage
{\bf Figures}

Figure 1: Powder X-ray (Cu $K_{\alpha}$ radiation) diffraction
spectra of stoichiometric MgB$_{2}$ (with $h,k,l$) for 3 different
boron qualities (a) 90\% pure natural boron; (b) 99.99\% pure
natural boron and (c) 99.95\% pure isotopic enriched $^{11}$B.
Samples (a) and (b) were synthesized for $3h/950^{\circ}C$, and
sample (c) for $4h/950^{\circ}C$. The data gaps are due to the
removal of the Si peaks.

Figure 2: Variation of the normalized zero-field resistance as a
function of temperature for MgB$_{2}$ pellets with different boron
purities. Inset: expanded scale near $T_{c}$.

Figure 3: Resistance curves normalized by their temperature
derivative at room temperature, for different boron purities.
Inset: expanded scale near $T_{c}$.

Figure 4: X-ray spectra for 3 different nominal compositions of
Mg$_{x}$$^{11}$B$_{2}$ for $x =$ 0.8, 1.0, 1.2.

Figure 5: Temperature dependence of the normalized resistance for
representative samples with nominal composition
Mg$_{x}$$^{11}$B$_{2}$ ($0.8 < x < 1.2$). Inset: expanded scale
near $T_{c}$.

Figure 6: Residual resistance ratio of Mg$_{x}$$^{11}$B$_{2}$
($0.8 < x < 1.2$). The open symbols represent different pieces
selected from the same batch. The solid symbols are the average.
The dotted box delimits the smaller variation ($x=1\pm 0.1$).

Figure 7: Resistance curves normalized by temperature derivative
at room temperature, for Mg$_{x}$$^{11}$B$_{2}$ ($x =$ 0.8, 0.9,
1.0, 1.1 and 1.2). $x=1.2$ data shown as dashed curve as discussed
in the text. Inset: expanded scale near $T_{c}$.

Figure 8: Powder X-ray  diffraction spectra of
Mg(B$_{0.8}$C$_{0.2}$)$_{2}$ (with $h,k,l$) for samples that were
synthesized for (a) 2 hours at $600^{\circ}C$ and then 2 hours at
$700^{\circ}C$; (b) 24 hours at $750^{\circ}C$; (c) 24 hours at
$950^{\circ}C$; (d) 24 hours at $1100^{\circ}C$ and (d) 24 hours
at $1200^{\circ}C$. The data gaps are due to the removal of the Si
peaks. Symbols: $+ =$ B$_4$C, $\ast =$ MgB$_2$C$_2$ and $\# =$
Mg$_2$C$_3$, as indicated in figure.

Figure 9: Temperature dependent magnetic susceptibility for
representative Mg(B$_{0.8}$C$_{0.2}$)$_{2}$ samples taken in $H =
50 Oe$ applied field: ZFC - warming. Inset: expanded temperature
range near $T_{c}$.

Figure 10: Normalized resistance as a function of temperature for
Mg(B$_{0.8}$C$_{0.2}$)$_{2}$ samples synthesized at different
reaction temperatures. Inset: expanded temperature range near
$T_{c}$.

\newpage
Table 1
\begin{center}
\small
\begin{tabular}{|c|c|c|c|}
\hline
  \bf{Purity} & \bf{Form} & \bf{Source} & \bf{Main Impurities}  \\\hline
  90\%   & \begin{tabular}{c}
     Amorphous \\
    (325 mesh) \\
  \end{tabular} & Alfa Aesar & \begin{tabular}{c}
     Mg \\
    5\%\\
  \end{tabular}\\\hline
  95\%   &  \begin{tabular}{c}
     Amorphous \\
    ($<$ 5 mesh) \\
  \end{tabular} &Alfa Aesar  &\begin{tabular}{c}
     Mg \\
    1\%\\
  \end{tabular}\\\hline
  98\%   & \begin{tabular}{c}
     Crystalline  \\
    (325 mesh) \\
  \end{tabular} & Alfa Aesar  &\begin{tabular}{c}
     C \\
    0.55\%\\
  \end{tabular}\\\hline
 99.95\%& \begin{tabular}{c}
     Isotopically pure $^{11}$B \\
    Crystalline (325 mesh) \\
  \end{tabular} & Eagle-Picher &\begin{tabular}{c}
     Si\\
    0.04\%\\
  \end{tabular}\\\hline
 99.99\% & \begin{tabular}{c}
     Amorphous \\
    (325 mesh) \\
  \end{tabular} & Alfa Aesar  & \begin{tabular}{c}
     Metallic Impurities \\
    0.005\%\\
   \end{tabular}\\ \hline
\end{tabular}
\end{center}

\newpage
\begin{figure}[htb]
Figure 1
\begin{center}
\includegraphics[angle=0,width=160mm]{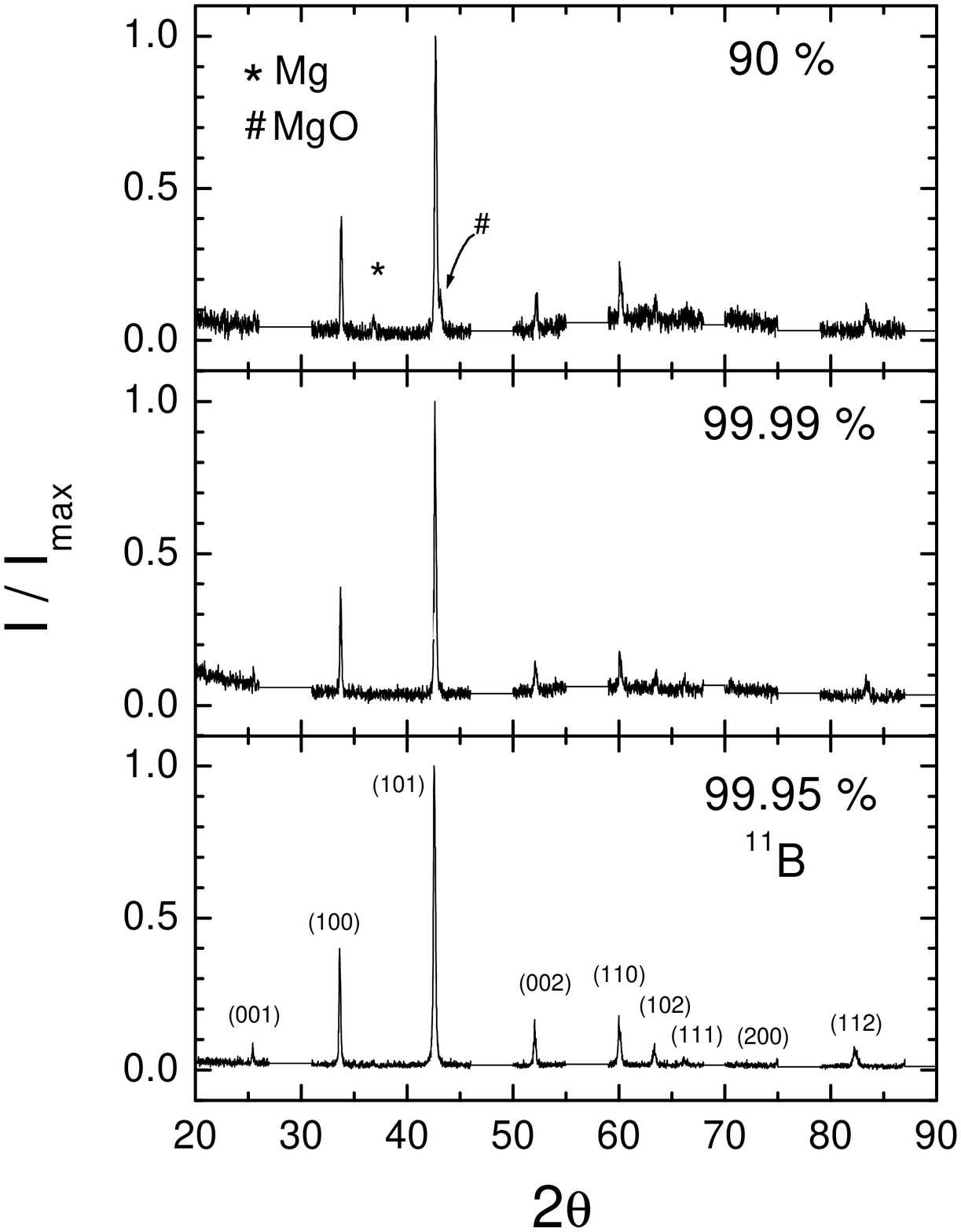}
\end{center}
\end{figure}

\begin{figure}[htb]
Figure 2
\begin{center}
\includegraphics[angle=0,width=160mm]{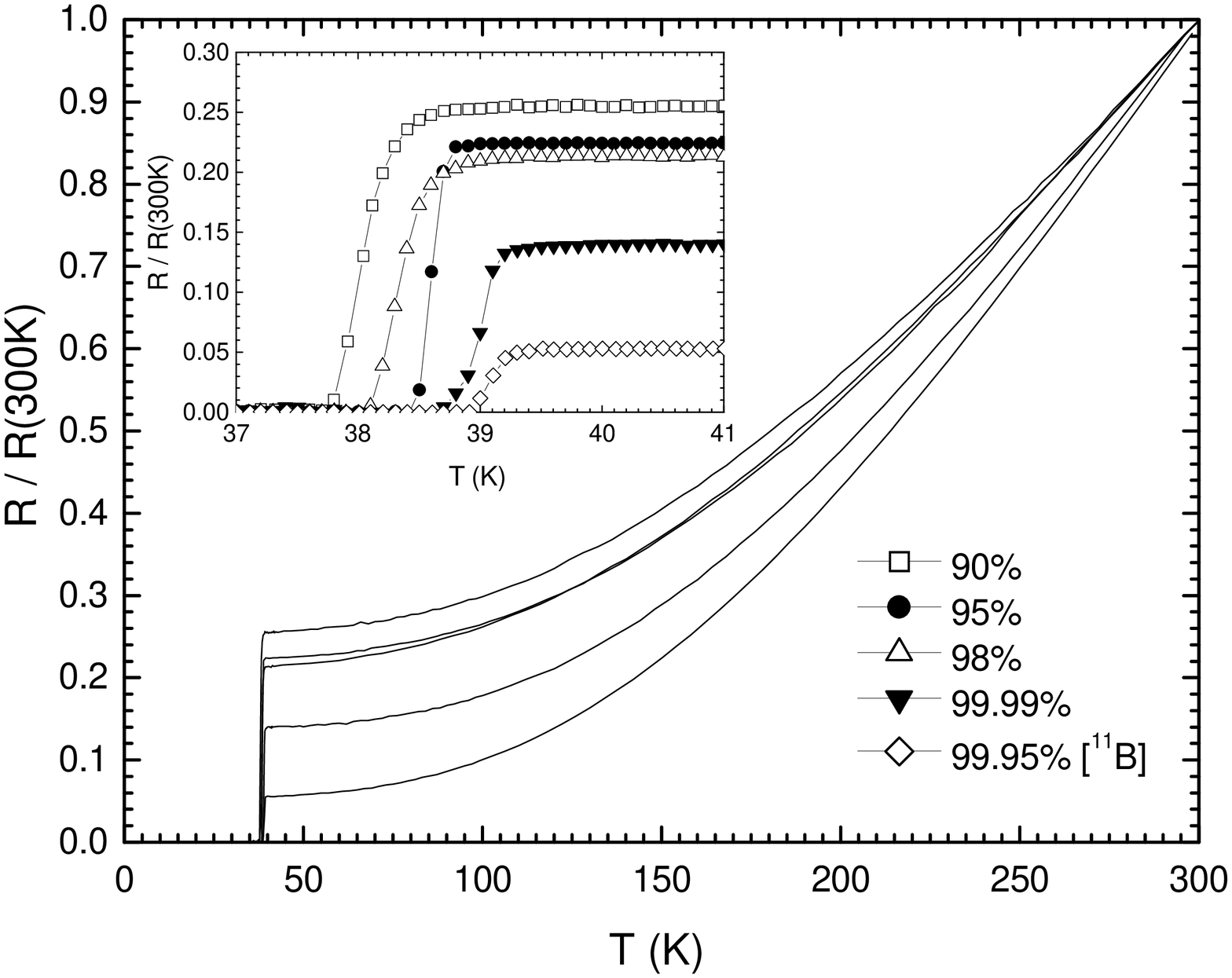}
\end{center}
\end{figure}

\begin{figure}[htb]
Figure 3
\begin{center}
\includegraphics[angle=0,width=160mm]{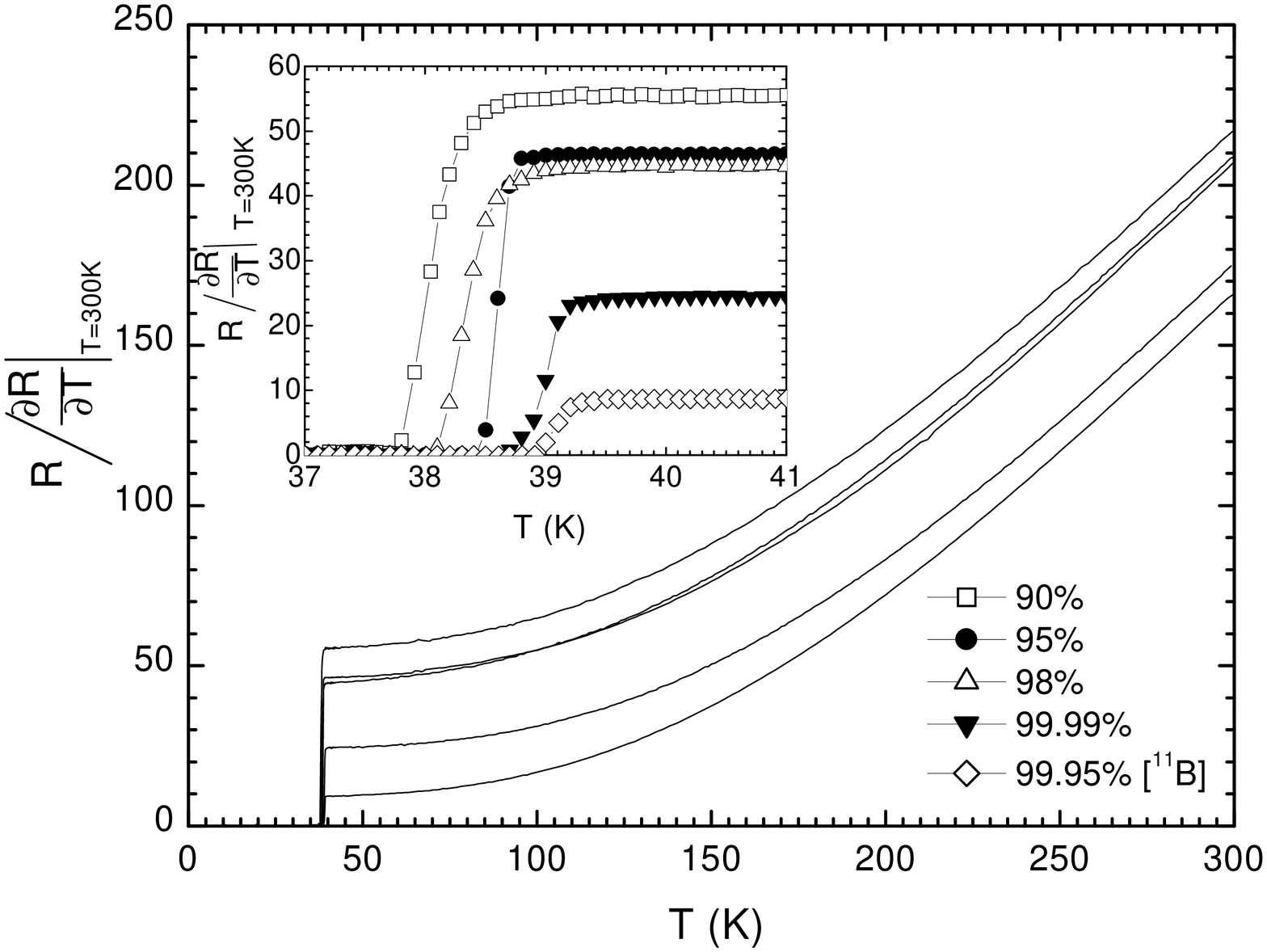}
\end{center}
\end{figure}

\begin{figure}[htb]
Figure 4
\begin{center}
\includegraphics[angle=0,width=160mm]{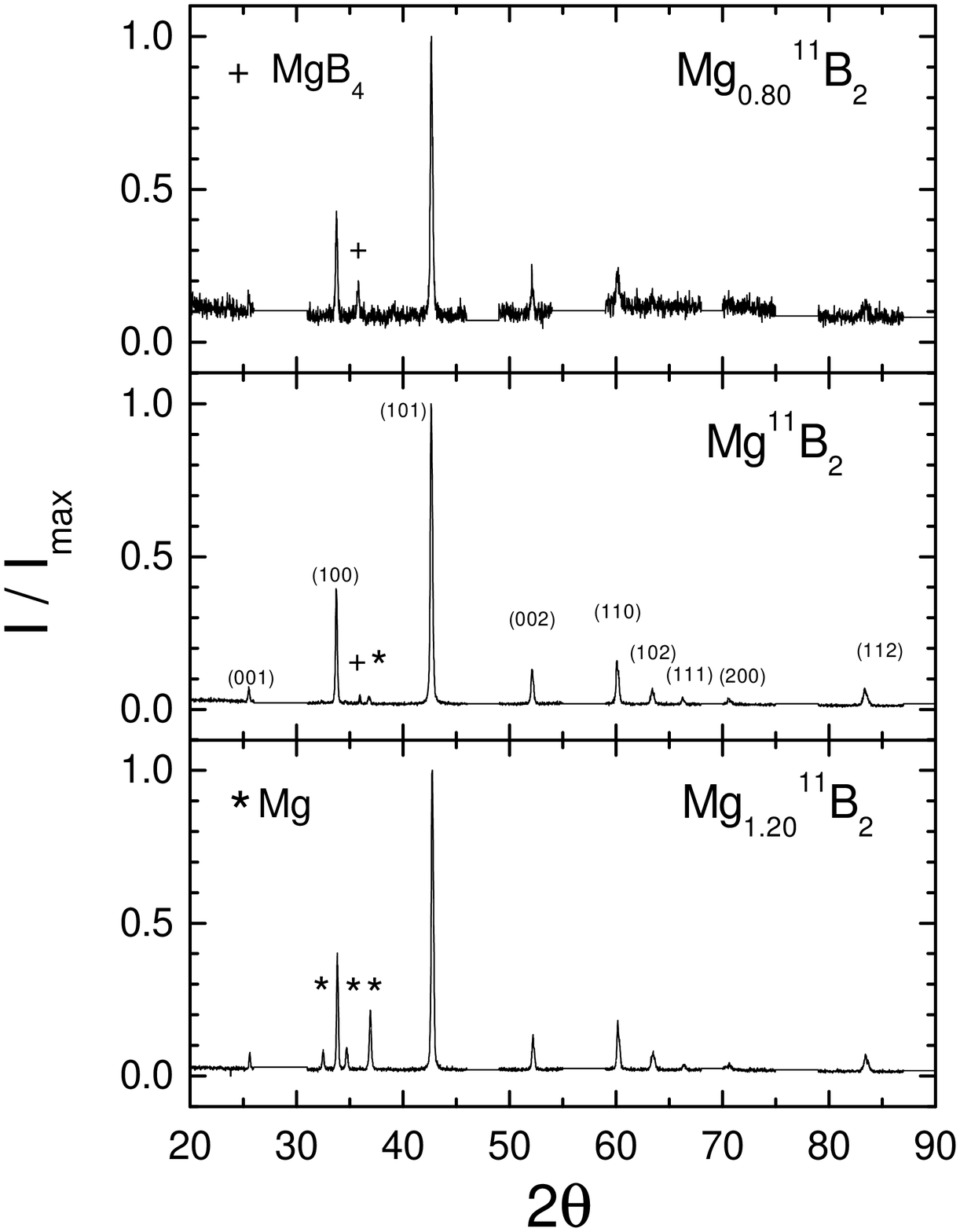}
\end{center}
\end{figure}

\begin{figure}[htb]
Figure 5
\begin{center}
\includegraphics[angle=0,width=160mm]{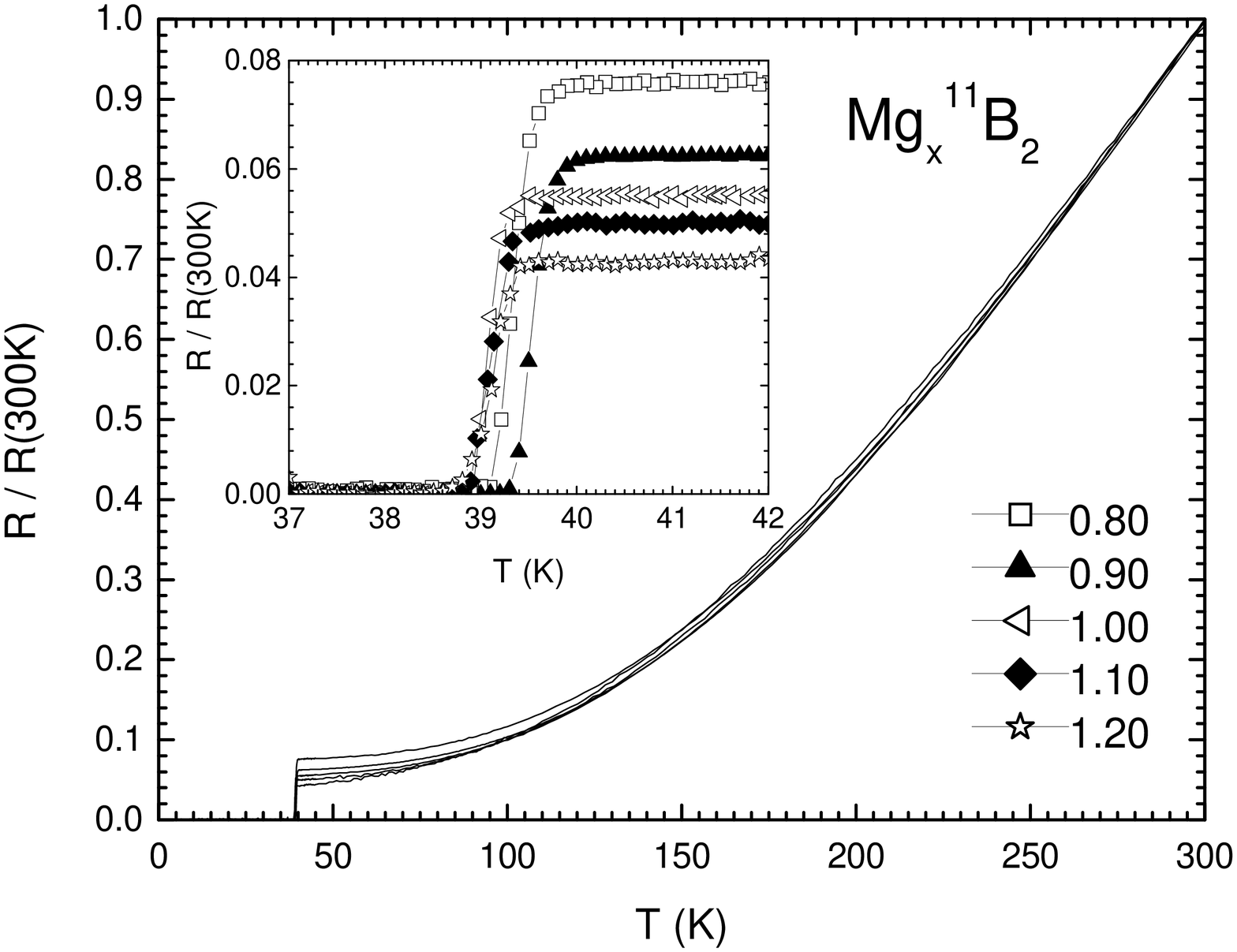}
\end{center}
\end{figure}

\begin{figure}[htb]
Figure 6
\begin{center}
\includegraphics[angle=0,width=160mm]{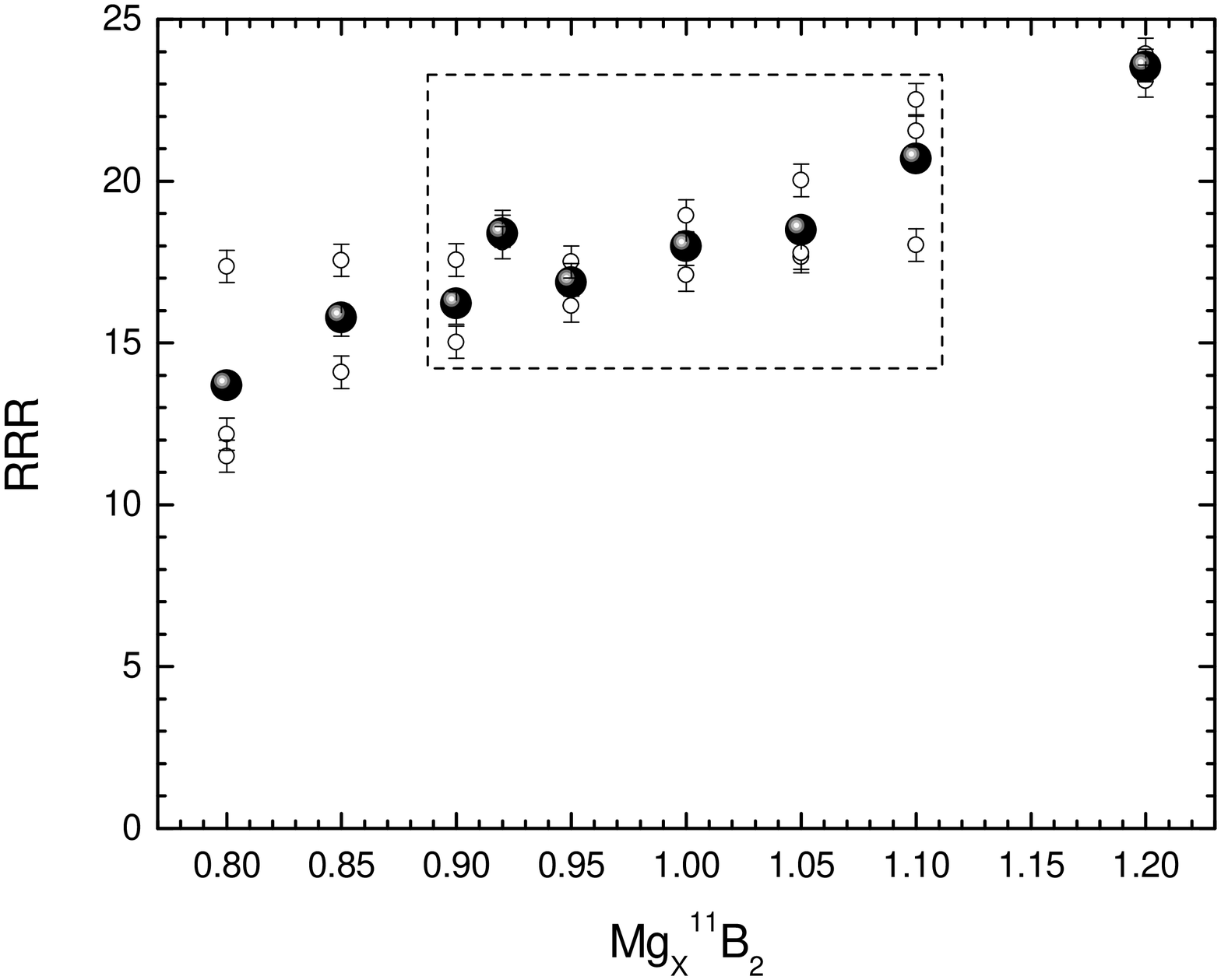}
\end{center}
\end{figure}

\begin{figure}[htb]
Figure 7
\begin{center}
\includegraphics[angle=0,width=160mm]{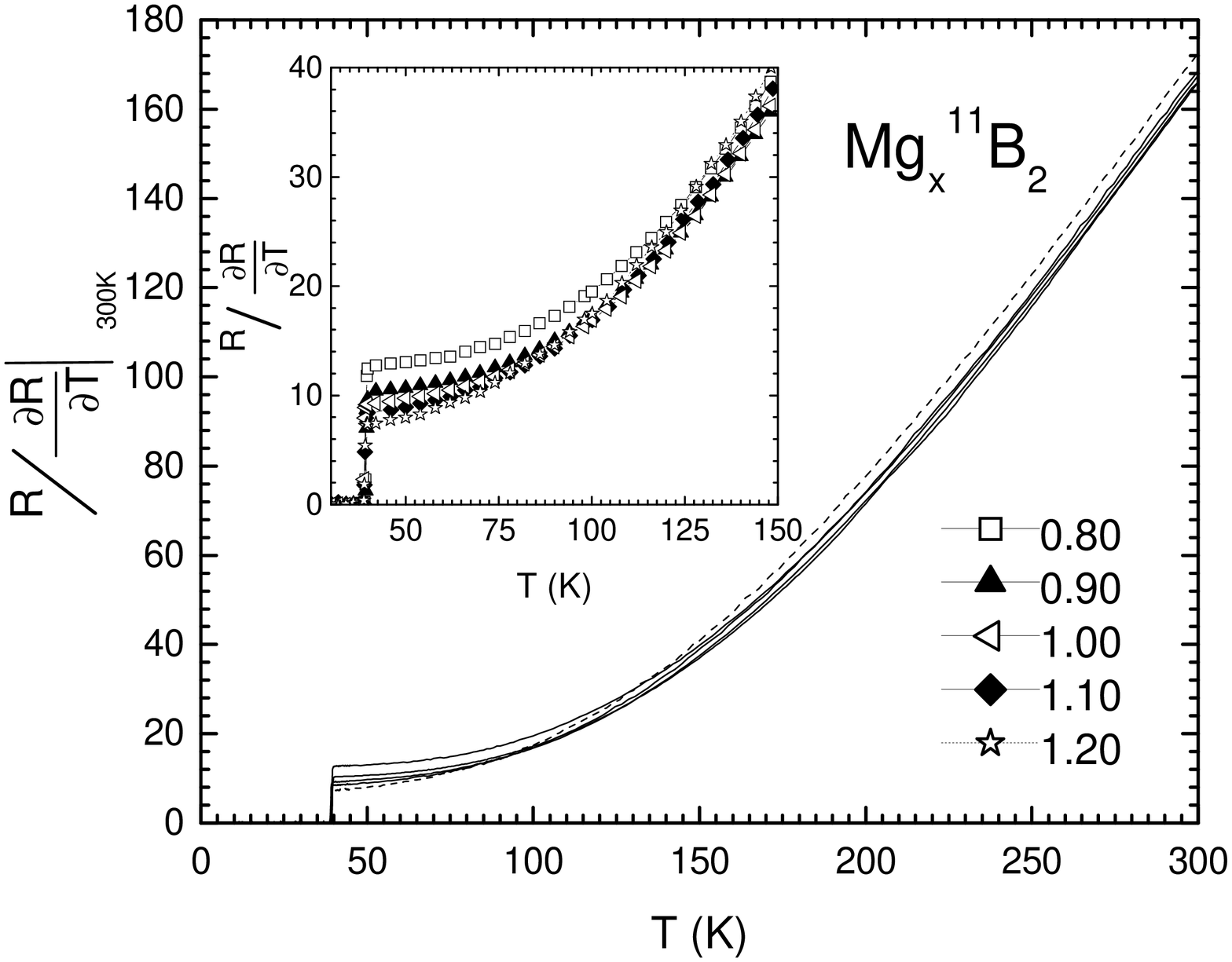}
\end{center}
\end{figure}

\begin{figure}[htb]
Figure 8
\begin{center}
\includegraphics[angle=0,width=160mm]{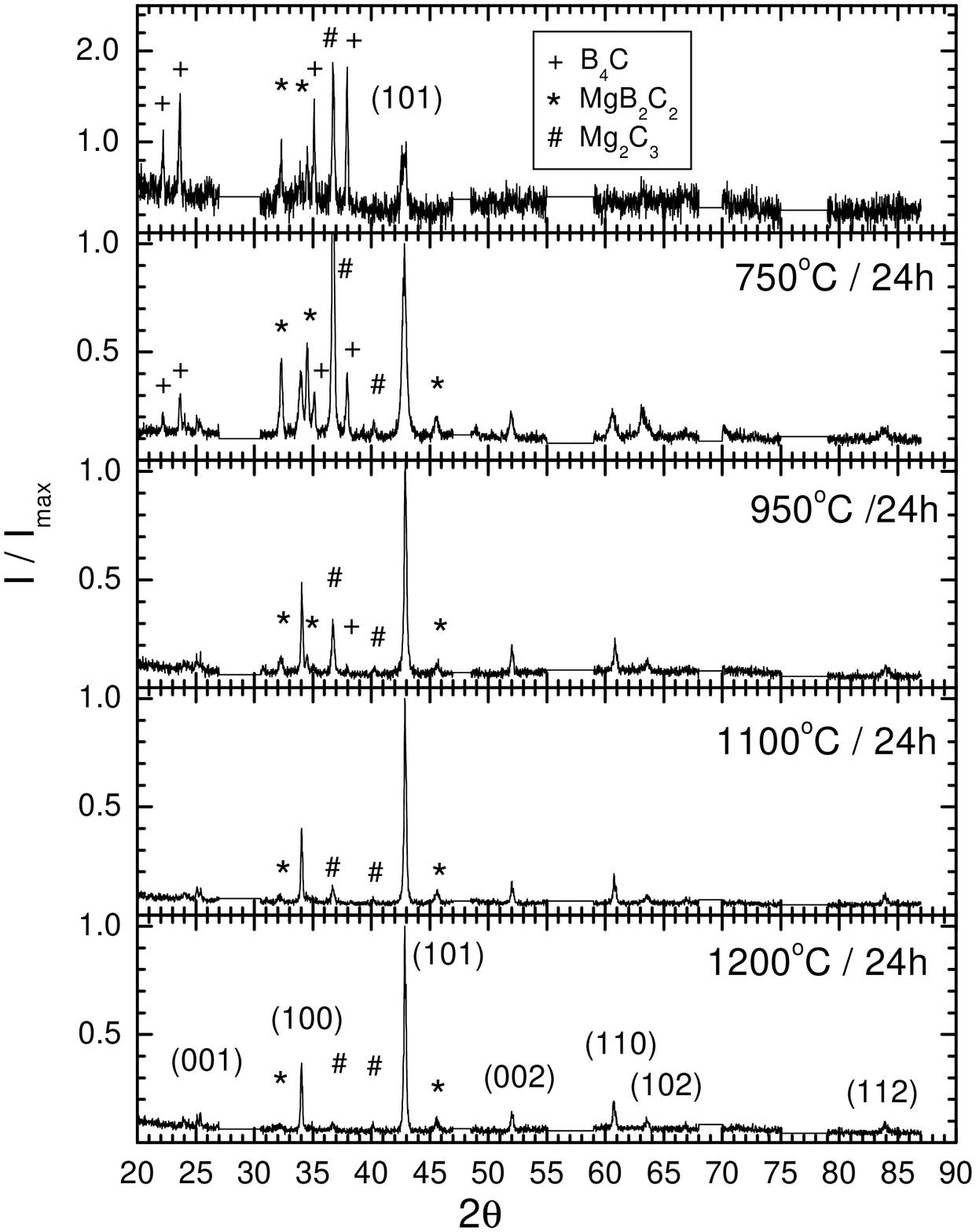}
\end{center}
\end{figure}

\begin{figure}[htb]
Figure 9
\begin{center}
\includegraphics[angle=0,width=160mm]{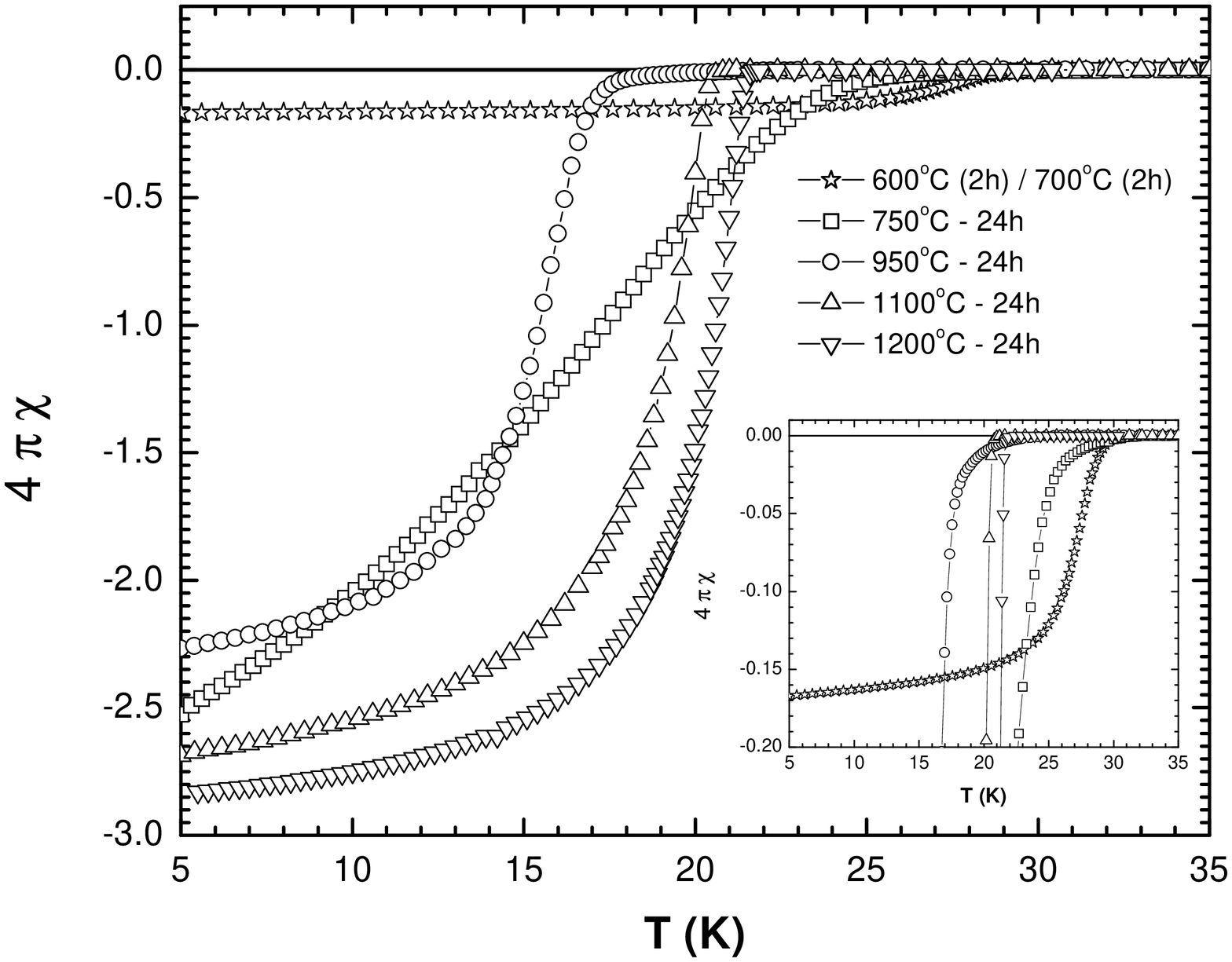}
\end{center}
\end{figure}

\begin{figure}[htb]
Figure 10
\begin{center}
\includegraphics[angle=0,width=160mm]{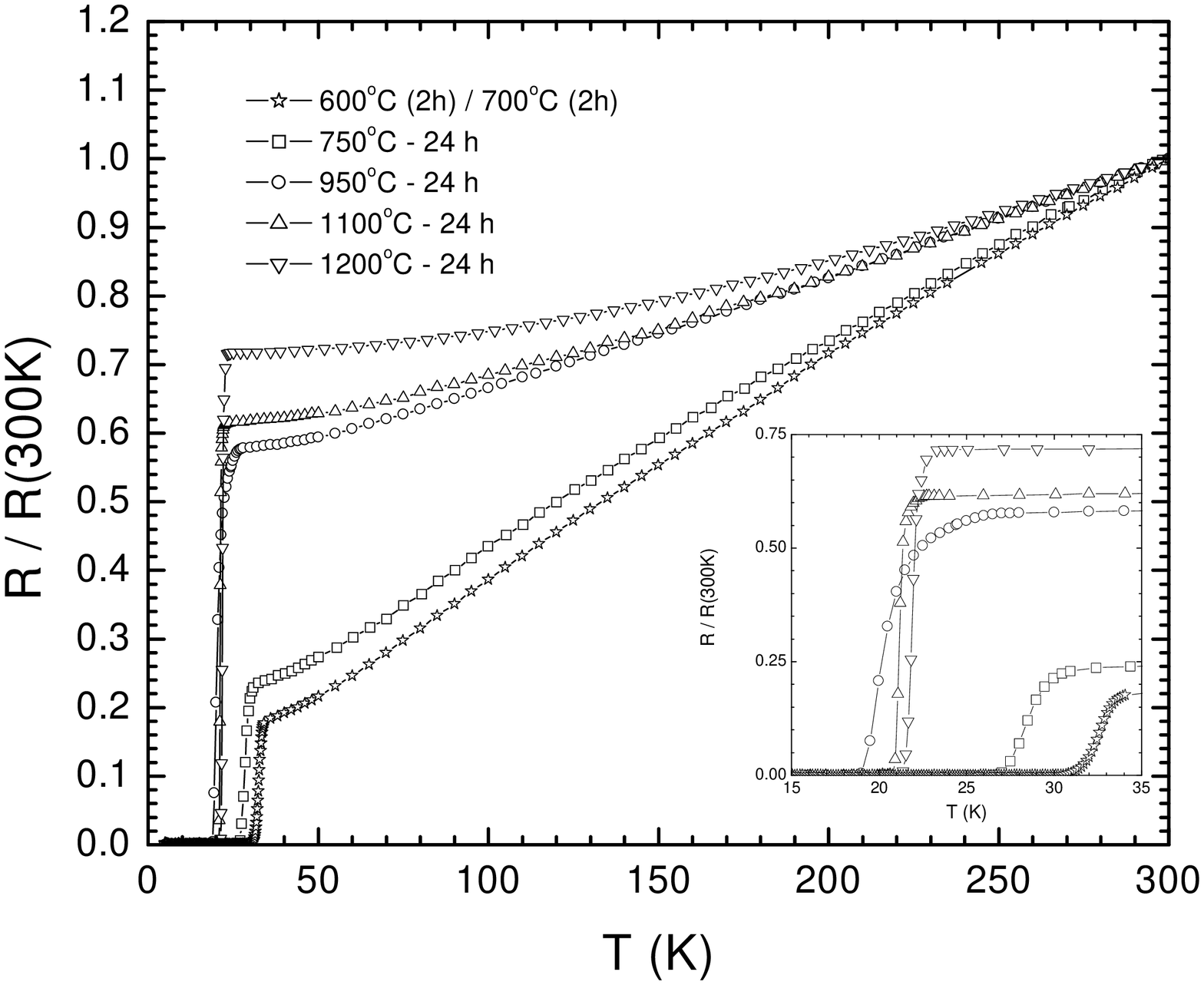}
\end{center}
\end{figure}

\end{document}